\documentclass[10pt, conference, compsocconf]{IEEEtran}
\usepackage[utf8]{inputenc}
\usepackage{graphicx}
\usepackage{cancel}
\usepackage{caption}
\usepackage{subfigure}
\usepackage{amsmath}
\usepackage{amsthm}
\usepackage{verbatim}
\usepackage[linesnumbered,ruled,vlined]{algorithm2e}

\usepackage{color} 

\newcommand{\striderlm}{Strider\textsuperscript{lsa} }

\newcommand{\ie}{\textit{i}.\textit{e}., }
\newcommand{\eg}{\textit{e}.\textit{g}., }

\newcommand{\sameAs}{\texttt{owl:sameAs} }

\newcommand{\remove}[1]{}

\begin{document}
\title{\striderlm: Massive RDF Stream Reasoning in the Cloud}
\author{\IEEEauthorblockN{Xiangnan Ren$^{1,2}$, Olivier Cur\'e$^{2}$ }
\IEEEauthorblockA{$^{1}$ATOS, 80 quai Voltaire, 95870 Bezons, France.\\$^{2}$LIGM (UMR 8049) CNRS, ENPC, ESIEE, UPEM,\\ Marne-la-Vall\'ee, France.\\
Email: \{firstname.lastname\}@u-pem.fr}
\and
\IEEEauthorblockN{Hubert Naacke$^{3}$, Ke Li$^{1,3}$}
\IEEEauthorblockA{$^{3}$LIP6 CNRS UMR 7606\\
Sorbonne Universit\'es, UPMC\\ Univ Paris 06, F-75005, Paris, France\\
Email: \{firstname.lastname\}@lip6.fr}
}

\maketitle

\begin{abstract} 
Reasoning over semantically annotated data is an emerging trend in stream processing aiming to produce sound and complete answers to a set of continuous queries. It usually comes at the cost of finding a trade-off between data throughput and the cost of expressive inferences. \striderlm proposes such a trade-off and combines a scalable RDF stream processing engine  with an efficient reasoning system. The main reasoning tasks are based on a query rewriting approach for SPARQL that benefits from an intelligent encoding of RDFS+ (RDFS + owl:sameAs) ontology elements. \striderlm runs in production at a major international water management company to detect anomalies from sensor streams. The system is evaluated along different dimensions and over multiple datasets to emphasize its performance.

\end{abstract}


\section{Introduction}
This paper deals with an emerging problem in the design of Big data applications: reasoning over large volumes of semantically annotated data streams. The main goal amounts to producing sound and complete answers to a set of continuous queries. This problem is quite important for many Big data applications in domains such as science, social and Internet of Things (IoT) in general. For instance, in the Waves\footnote{http://www.waves-rsp.org/} project, we are dealing with ``real-time" anomaly detection in large water distribution networks. By working with domain experts, we found out that such detections can only be performed using reasoning services.

Tackling this issue implies to find a trade-off between data throughput and reasoning over semantically annotated data streams. This is notoriously hard and although it is currently getting some attention, it is still an open problem. RDF Stream Processing (RSP) engines are the prominent systems tackling this problem where annotation are using the RDF (Resource Description Framework)\footnote{https://www.w3.org/RDF/} data model, queries are expressed in a continuous SPARQL form and reasoning are supported by RDFS/OWL ontologies. Existing RSP engines are either not scalable (\ie they do not distribute data and/or processing) or do not support expressive reasoning services.

Our system, \striderlm, combines our Strider RSP engine with a reasoning approach, denoted LiteMat. Strider\cite{DBLP:journals/corr/RenC17} is a stream processing engine for semantic data taking the form of RDF  graphs. It is designed on top of state-of-the-art Big Data components such as Apache Kafka and Apache Spark. It is thus the first RSP engine capable of handling at scale high throughput with relatively low latency. Strider is capable of processing and adaptively optimizing continuous SPARQL queries. Nevertheless, it was not originally designed to perform inferences. Hence, the motivation is to integrate an extension of our RDFS LiteMat\cite{DBLP:conf/bigdataconf/CureNRA15} inference  approach. Intuitively, it provides a trade-off between the most common reasoning methods encountered in  Knowledge Bases (KB): query rewriting and materialization.

The LiteMat extension we are considering in this work integrates the popular, \eg in Linked Open Data (LOD) KBs, \sameAs property. Intuitively, this property enables to define aliases between RDF resources. This is frequently used when a domain's (meta)data is described in a collaborative way, \ie a given object has been given different identifiers (possibly by different persons) and are later reconciled by stating their equivalence. We discovered several of these situations in the context of the Internet of Things (IoT) Waves project. For instance, we found out that sensors or locations in water networks could be given different identifiers.
Generally, the triples containing a \sameAs property are not present in the streaming data but are rather stored in static metadata, \eg a RDF store\cite{Cure:2014:RDS:2785631}. Such metadata are needed to perform valuable inferences. In the Waves project, they correspond to the topology of the network, characteristics of the network's sensors, etc. We consider that the presence of static metadata can be generalized to many domains, \eg life science, social, cultural, and it is hence important to design a solution to reason over their data streams.

The main contributions of this paper are (i) to combine a scalable, production-ready RSP engine that supports reasoning services over RDFS+ (\ie RDFS + \sameAs) KB, (ii) to minimize the reasoning cost, and thus to guarantee high throughput and acceptable latency, and (iii) to propose a thorough evaluation of the system and thus to highlight its relevance.

\section{Background knowledge}
\subsection{RDF, SPARQL, RDFS and OWL}
Data on the Web is generally represented using RDF, a schema-free data model.
Assuming disjoint infinite sets I (RDF IRI references), B (blank nodes) and L (literals), a triple (s,p,o) $\in$ (I $\cup$ B) x I x (I $\cup$ B $\cup$ L) is called an RDF triple with s, p and o 
respectively being the subject, predicate and object. 
We now also assume that V is an infinite set of variables and that it is disjoint with I, B and L. 
%
SPARQL is the W3C query language recommendation for the RDF format.
We can recursively define a SPARQL\cite{sparql11} triple pattern (tp) as follows: 
(i) a triple $tp \in$ (I $\cup$ V) x (I $\cup$ V) x (I $\cup$ V $\cup$ L) is a SPARQL triple pattern, (ii) if $tp_1$ and $tp_2$ are triple patterns, then ($tp_1 . tp_2$) represents a group of triple patterns 
that must all match, ($tp_1$ \texttt{OPTIONAL} $tp_2$) where $tp_2$ is a set of patterns that may extend the solution induced by $tp_1$, and ($tp_1$ \texttt{UNION} $tp_2$), denoting pattern alternatives, are 
triple patterns and (iii) if $tp$ is a triple pattern and C is a built-in condition, then, ($tp$  \texttt{FILTER} C) is a triple pattern enabling to restrict the solutions of a triple pattern match according to the expression C. A set of tps is denoted a Basic Graph Pattern (BGP). The SPARQL syntax follows the select-from-where approach of SQL queries. 

We consider that a KB consists of an ontology, aka terminological box (Tbox), and a fact base, aka assertional box (Abox). The least expressive ontology language of the Semantic Web is RDF Schema\cite{rdfs11} (RDFS). Its goal is to provide a mechanism allowing to describe groups of related resources (concepts) and their relationships (properties).
RDFS entailment can be computed using 14 rules. But practical inferences can be computed with a subset of them. 
The one we are using is $\rho$df which has been defined and theoretically investigated in \cite{Munoz:2007:MDS:1419662.1419670}. In a nutshell, $\rho df$ considers inferences using \texttt{rdfs:subClassOf}, \texttt{rdfs:subPropertyOf}, \texttt{rdfs:range} and \texttt{rdfs:domain} properties.
An RDF property is defined as a relation between subject and object resources. RDFS allows to describe this relations in terms of the classes of resources to which they apply by specifying the class of the subject (i.e., the domain) and the class of the object (i.e., the range) of the corresponding predicate. The corresponding \texttt{rdfs:range} and \texttt{rdfs:domain} properties allow to state that respectively the subject and the object of a given \texttt{rdf:Property} should be an instance of a given \texttt{rdfs:Class}.
The property \texttt{rdfs:subClassOf} is used to state that a given class (i.e., \texttt{rdfs:Class}) is a subclass of another class. 
Similarly, using the property \texttt{rdfs:subPropertyOf}, one can state that any pair of resources (i.e., subject and object) related by a given property  is also related by another property.

Other ontology languages, OWL\footnote{https://www.w3.org/TR/owl2-overview/} (Web Ontology Language) an its fragments, of the Semantic Web stack are more expressive than RDFS, \eg supporting \sameAs. The deductive process has a higher computational cost which can become a problem if low latency is expected.


Two main approaches are generally used to support inferences in KBs. The first approach
consists in materializing all derivable triples before evaluating any queries.  It implies a possibly long loading time due to running reasoning services during a data preprocessing phase. This generally drastically increases the size of the buffered data and imposes specific dynamic inference strategies when data is updated. Besides, data materialization also potentially increases the complexity for query evaluation (\eg longer processing for table scanning). These behaviors can seriously impact query performance. The second approach consists of reformulating each submitted query into an extended one including semantic relationships from the ontologies. Query rewriting avoids costly data preprocessing, storage extension and complex update strategies but induces slow query response times since all the reasoning tasks are part of a complex query preprocessing step. 

In a streaming context, due to the possibly long lifetime of continuous queries, the cost of query rewriting can be amortized.
On the other hand, materialization tasks have to be performed on each incoming streams, possibly on rather similar sets of data, which implies a high processing cost, \ie lower throughput and higher latency.

\subsection{RDF Stream Processing (RSP)}
The nature of stream processing is to run a set of operations over unbounded data streams. Such processing are generally performed within a windowing operator that slices the incoming infinite data streams into finite chunks. These windows can be defined over some temporal constraints, \eg take the last 3 minutes of incoming data, or over non-temporal parameters, \eg counting or session based.

In the last decade, much effort has been devoted to improving RSP. Systems like C-SPARQL \cite{CSPARQL} , CQELS \cite{CQELS}, SparqlStream \cite{SPARQLStream} and ETALIS \cite{ETALIS} provide a SPARQL-based continuous query language to cope with RDF data streams. The aforementioned RSP systems give a basic solution to perform SPARQL query over RDF data stream.
Whilst, C-SPARQL and SparqlStream tackle the requirements of stream reasoning via data materialization and query rewriting, respectively. The proposed straightforward solutions, \ie data materialization and query rewriting show limitations on systems' scalability. Moreover the centralized design of C-SPARQL and SparqlStream can not support complex reasoning task over massive RDF data stream. On the other hand, distributed RSP engines, \eg CQELS-Cloud, address flexibility and scalability issues but do not possess of real-time stream reasoning.

In summary, although RSP engines have substantially improved in recent years, none of them cover the scalability and expressive reasoning aspects.

\section{\striderlm overview}
\begin{figure}[ht]
\vspace{-2mm}
\advance\leftskip-0.3cm
\includegraphics[scale=0.4]{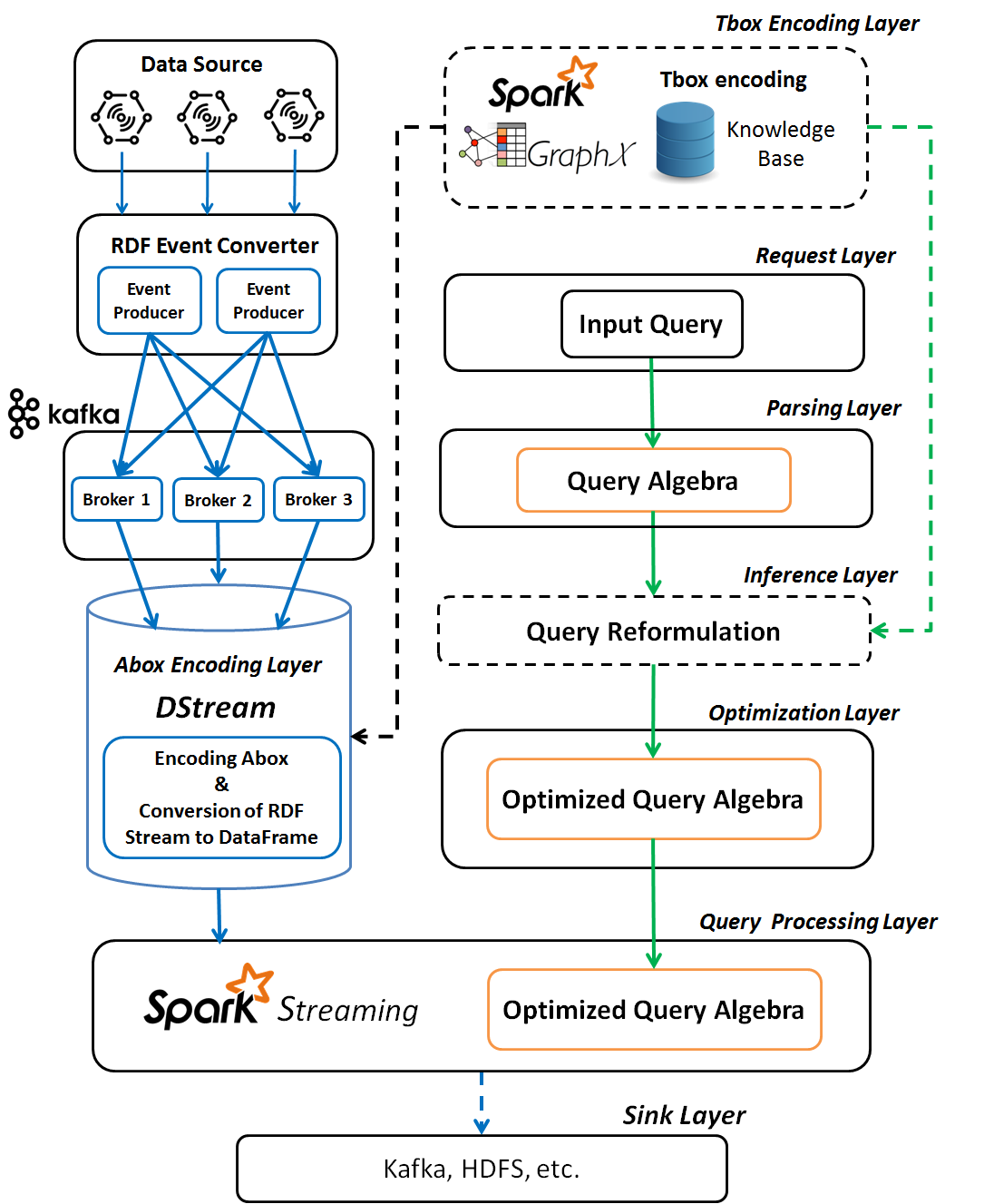}
\caption{\striderlm Architecture}
\label{fig:architecture}
\end{figure}

In this section, we first give a high-level view on the architecture of \striderlm (see Figure \ref{fig:architecture}).

\striderlm consists of two principal modules: (1) data flow management. For the purpose of ensuring high throughput and fault-tolerance, we use Apache Kafka to manage the data flow of \striderlm. The incoming data are assigned to  Kafka \emph{topics}, and partitioned among a cluster to enable parallelism of upstream/downstream operations. (2) Computing core. The query processing pipeline is implemented in the Apache Spark programming framework. Spark Streaming continuously receives data from Kafka, and performs SPARQL queries executions in parallel. We briefly illustrate the general process of query processing and stream reasoning in \striderlm as follow: With a given ontology, the Tbox encoding layer runs a mostly distributed off-line knowledge base encoding for further reasoning request. Once a query is registered, the system will, if necessary, first rewrite the query into a LiteMat (see \ref{subsec:litemat} for more details) representation and construct the corresponding query logical plan. Then, \striderlm pushes the obtained logical plan into the query processing layer for a continuous SPARQL query evaluation. Note that, \striderlm is capable of adjusting the query execution plan at-runtime via its optimization layer.

\section{Reasoning in \striderlm}




%
%
%
%
%













\subsection{Encoding}
\label{subsec:litemat}

\subsubsection{Static TBox encoding}
\label{stbox}
\textbf{Concept and property hierarchies  encoding} is needed upfront to any data stream processing. The TBox Encoding Layer encodes concepts, properties and instances of registered KBs.  This aims to provide efficient encoding scheme and data structures to support the reasoning services associated to the input ontology of an application. 
The input ontology is considered to be the union of (supposedly aligned) ontologies necessary to operate over one's application domain. 

In the following, we consider the $\rho$df subset of RDFS and use our LiteMat approach\cite{DBLP:conf/bigdataconf/CureNRA15}. To address inferences drawn from \texttt{rdfs:subClassOf}, \texttt{rdfs:subPropertyOf} , we attribute numerical identifiers to ontology terms, \ie concepts and properties. 
The compression principle of this term encoding lies in the fact that subsumption relationships are represented within the encoding of each term. 
This is performed by prefixing the encoding of a term with the encoding of its direct parent (a workaround is proposed to support multiple inheritance). 
This approach only works if an encoding is computed using a  binary representation.  

More precisely, the concept (resp. property) encoding are performed in a top down manner, \ie starting from the top concept of the hierarchy (the classification is performed by a state-of-the-art reasoner, \eg HermiT\cite{hermit}, and hence supports all OWL2 logical concept subsumptions), such that the prefix of any given sub-concept (resp. sub-property) corresponds to its super-concept (resp. super-property). Since, the lengths of concept (resp. property) hierarchy branches may be  different, the system normalizes all identifiers to ensure that they are of the same length. The characteristics of this encoding scheme ensures that from any concept (resp. property) element, all its direct and indirect sub-elements can be computed with only two  bit shift operations and are comprised into a discrete interval of integer values, namely its lower and upper bound (resp. LB,UB). Table \ref{table:ssn} presents the identifiers of the encoding of the Semantic Sensor Network (SSN) concept hierarchy. We can observe that the \texttt{Action}'s prefix 111110 corresponds to \texttt{Event}'s identifier, and his hence one of its direct sub-concept. Moreover, \texttt{Event} is a direct sub-concept of \texttt{Entity} and indirect sub-concept of \texttt{owl:Thing}.
More details on the encoding scheme can be obtained in \cite{DBLP:conf/bigdataconf/CureNRA15}.

\begin{table}[!h]
\centering 
\begin{tabular}{|l|c|l|}
\hline
Raw ids & Normalized ids & Term\\
\hline
1 & 1000000000 & \texttt{owl:Thing}\\
\hline
101 & 1010000000 & Input\\
\hline
110 & 1100000000 & Output\\
\hline
111 & 1110000000 & Entity\\
\hline
111001 & 1110010000 & Abstract\\
\hline
111010 & 1110100000 & FeatureOfInterest\\
\hline
111011 & 1110110000 & InformationEntity\\
\hline
111100 & 1111000000 & Object\\
\hline
111101& 1111010000& Quality\\
\hline
111110 & 1111100000 & Event\\
\hline
1111100010 & 1111100010 & Action\\
\hline
\end{tabular}
\caption{Encoding for an extract of the concept hierarchy of the SSN ontology}
\label{table:ssn}
\vspace{-5mm}
\end{table} 

\textbf{Individual encoding} can take two different forms, depending on whether an individual is involved in a triple with a \sameAs property or not. For non-sameAs individuals, we apply a simple method which attributes a unique integer identifier (starting from 1) to each individual. In \cite{DBLP:conf/bigdataconf/CureNRA15}, we provided an efficient distributed method to perform this encoding.

\begin{figure}[ht]
\includegraphics[scale=0.38]{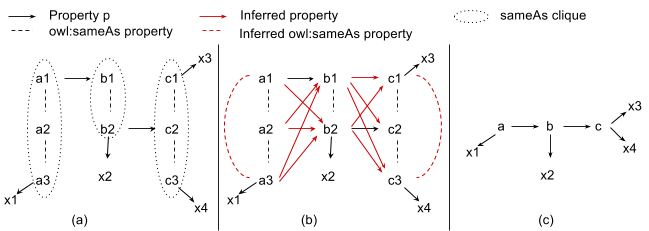}
\caption{\sameAs representation solutions}
\label{fig:sameAsRep}
\end{figure}

The support for individual involved in triples with \sameAs properties (denoted sameA has been integrated in  our LiteMat approach in the context of this work. It is motivated by the popularity of \sameAs across domains of the LOD. For instance, the \sameAs constructor is frequently encountered to practically define or maintain ontologies. In \cite{Halpin10whenowl:sameas}, the authors measured the frequency of \sameAs~in an important repository of LOD. That property was involved in 58,691,520 triples over 1,202 unique domain names with the most popular domains being biology, \eg bio2rdf and uniprot (respectively 26 and 6 million triples involving \sameAs), and general domains, \eg DBpedia (4.3 million \sameAs~triples).
Moreover, the knowledge management of LOD, estimated to more than 100 billion triples, clearly amounts to big data issues. In our Waves running example, we also found out that, due to the cooperative ontology building, RDF triples containing \sameAs properties were necessary to re-conciliate ontology designs.


In order to reason over KB supporting the \sameAs property, we can use different approaches. An obvious and naive one materializes all inferences. For instance, from Fig.\ref{fig:sameAsRep}(a), we obtain Fig.\ref{fig:sameAsRep}(b) where all red properties correspond to materialization, resulting in doubling the number of triples (without materializing the symmetric aspect of \sameAs).
Thus one of our challenge is to support sameAs reasoning while answering queries without materialization. In \striderlm,~we address this challenge with an approach consisting in selecting a single individual among a clique of sameAs individuals (in Fig. \ref{fig:sameAsRep}(a), three cliques are represent in dotted ellipses). This approach has many advantages, especially in a data streaming context: (i) the inferred graph is more compact without loss of information than the  original graph, (ii) with the proper approach and dictionary data structures, the computing overhead of transforming the graph is not issue and (iii) clique updates (\eg removing or adding an individual from a clique) has no performance impact since the data streams are ephemeral. Fig.\ref{fig:sameAsRep}(c) is an example of this approach where individuals a,b and c represent the so-called representatives of the cliques. 

Our sameAs encoding method aims to guarantee advantages (i) and (ii). This is performed by encoding individuals involved in a sameAs clique with a binary tuple identifier where the first value correspond to a clique identifier and the second one to a local identifier in that clique. 

In a nutshell, a static RDF dataset is parsed to extract all triples involving  a \sameAs property. From that graph, a connected component operation is computed to define all the cliques. Then each clique is attributed a unique  integer value and for each clique, each individual is also given a unique integer value. A dictionary, taking the form of a hash map, is created out of this processing: the key is the IRI or literal of the individual and the value is the tuple identifier. This dictionary can be reversed to translate query results. This processing is computed in a distributed and parallel manner using the Apache GraphX component. Hence it is able to scale to very large static KBs.

\begin{figure}[ht]
\advance\leftskip-0.3cm
\includegraphics[scale=0.45]{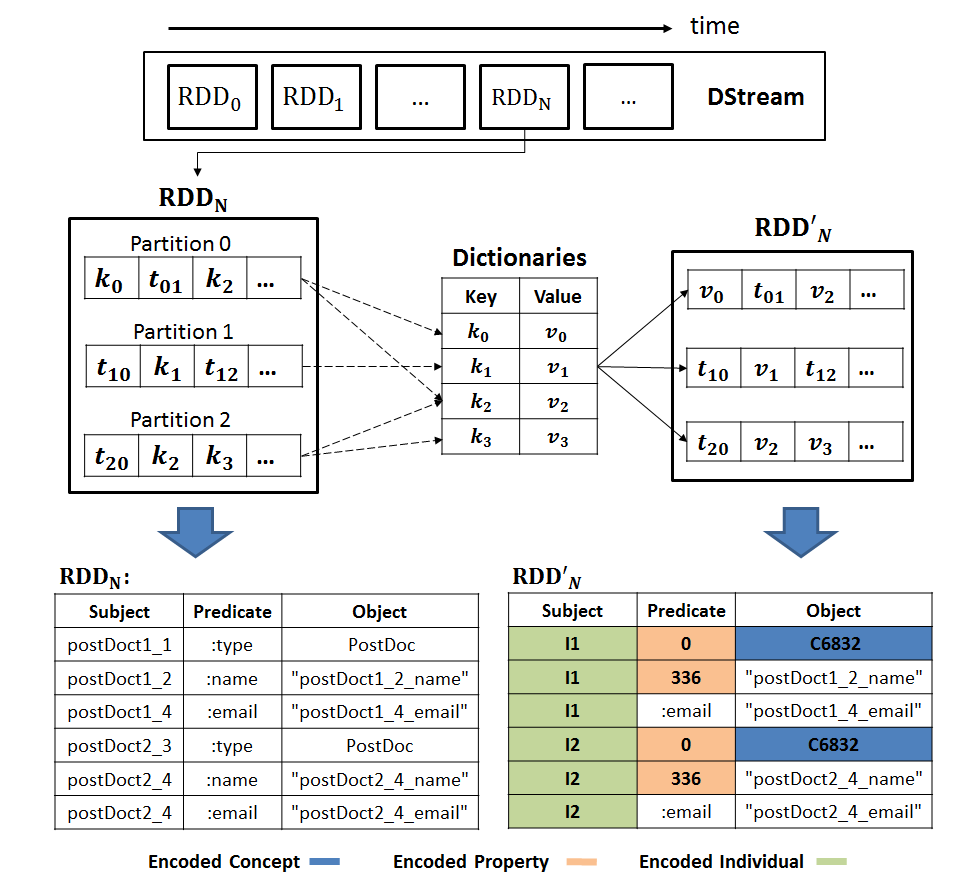}
\caption{Parallel partial encoding over DStream}
\label{fig:partialEncod}
\vspace{-3mm}
\end{figure}

\subsubsection{Dynamic Abox stream encoding}

In the previous section, we have encoded the elements of the static KB, \ie its concepts, properties and individuals. In this section, we explain how data streams are encoded on the fly and how it is mixing static (Tbox and Abox) and dynamic elements (Abox only since it is not probable that ontology modifications are streamed). In the following, we consider that concept, properties and individuals can or can not be in the precomputed dictionaries (nevertheless, note that the absence of concepts and properties in the dictionaries prevents from any reasoning tasks upon them). For instance, it makes sense that blank nodes in data streams are not in our individual dictionary while IRIs and literals may be present. In case of absence, the idea is not to encode the triple element since it would be too costly (generating a unique integer for a possibly infinite set of potential values) and it would possibly not be worth it since that identifier is essentially ephemeral. We thus qualify this encoding as a partial one and describe it in Fig.\ref{fig:partialEncod} which uses the Discretized stream(Dstreams) \cite{dstreams} abstraction of Apache Spark Streaming. Intuitively, a Dstreams is a set of RDD (Resilient Distributed Datasets)\cite{DBLP:journals/cacm/ZahariaXWDADMRV16} which are partitioned over a set of machines. Each RDD is composed of a set of RDF triples. For each RDD, a transformation is performed which takes a IRI/literal based representation to a partially encoded form. This transformation lookups into the TBox and Abox dictionaries precomputed from static KBs. The bottom right table of the figure emphasizes that some triple elements are encoded while some other are not. To apply our representative-based approach, each sameAs individual is replaced only with its clique identifier. Each non-sameAs individual is replaced with its corresponding integer identifier.

We briefly summarize important advantages of the partial encoding of RDF streams: (i) an efficient parallel encoding to meet real-time request; (ii) no extra overhead for dictionary maintenance; (iii) an off-line dictionary encoding on \texttt{owl:sameAs} with no extra overhead for materialization of data stream.

\subsection{Query Rewriting and Processing}
\label{qrew}
The query rewriting in \striderlm is done in the Inference Layer. Intuitively, the system parses a given SPARQL query Q and rewrites it into Q'. This rewriting concerns inferences pertaining to concept and property hierarchies. For sameAs individuals, no specific rewriting is necessary due to our data streams encoding. Due to space limitations, we do not present the rewriting of SPARQL queries into Spark SQL Scala programs but the interested reader can find details on our github page.

\subsubsection{Rewriting pertaining to hierarchies}
In Section \ref{stbox}, we highlighted that to each concept and property corresponds a unique integer identifier. Moreover, one characteristic of our encoding method guarantees that all sub-concept (resp. sub-property) identifiers of a given concept (resp. property) are included into an interval of integer values, denoted lower bound (LB) and upper bound (UB) of that ontology element.

We first present the rewriting for concepts. In order to speed up the rewriting, we take advantage of the following context: concepts are necessarily at the object position of a triple pattern and the property must be \texttt{rdf:type}. Intuitively, if a concept has at least one sub-concept then it is replaced in the triple pattern by a novel variable and a SPARQL \texttt{FILTER} clause is added to the query's BGP. That filter imposes that the new variable is included between the LB and UB values (which have been previously computed at encoding-time and stored in the dictionary) of that concept.

The overall approach is quite similar for the rewriting concerning the property hierarchy but no specific context applies, \ie all triple patterns have to be considered. For each triple pattern, we check whether the property has some sub-properties. It it is the case then the property is replaced by a new variable in the triple pattern and a SPARQL \texttt{FILTER} clause is introduced in the BGP. That filter clause restricts the new variable to be included in the LB and UB of that property.

As a concrete example of this rewriting, we are using query Q4 of our benchmark since it requires inferences over both the concept and property hierarchies.

\begin{center}
\begin{verbatim}
SELECT ?o ?n 
WHERE { ?x rdf:type lubm:Professor; 
        memberOf ?o; lubm:name ?n.}
\end{verbatim}
\end{center}

The rewriting Q4' of Q4 contains two \texttt{FILTER} clauses, one for the Professor concept and one for the memberOf property (LB() and UB() functions respectively return the LB and UB of their parameter): 

\begin{center}
\begin{verbatim}
 SELECT ?o ?n 
 WHERE { ?x rdf:type ?p; 
         ?m ?o; lubm:name ?n.
         FILTER (?p>=LB(Professor) 
             && ?p<UB(Professor)).
         FILTER (?m>=LB(memberOf) 
             && ?m<UB(memberOf)).}
\end{verbatim}
\end{center}

Note that this rewriting is much more compact and efficient than the classical reformulation which would require 8 \texttt{UNION} clauses and 18 joins\footnote{Rewriting available on our github page}

\subsubsection{Processing pertaining to sameAs cliques}
Recall that sameAs individuals are identified by a tuple of the form (cliqueId, localId) where cliqueId is unique for each sameAs clique and localId corresponds  to the local identifier of an individual in that clique. Moreover, data streams containing sameAs individuals are encoded with the cliqueId only, thus permitting to consider the whole clique.

Given this encoding, a standard query processing is performed where variable bindings concern both standard and sameAs individuals.

\subsubsection{Configuring the inference methodology}
\label{config}
We found out that in a streaming context, the ability to select which inference approach to execute may be quite useful. So far, we have presented our LiteMat-based approach but we have also introduced a full materialization approach. In Section \ref{eval}, we compare our liteMat approach to a SameAs Materialization (SAM) approach.  The configuration enables to select one method that the end-user may consider to be more efficient. Hence it is possible to configure a registered query to use a precise reasoning method. We have achieved this by extending our previous work \cite{DBLP:journals/corr/RenC17} on SPARQL-based query language with keywords configuring this feature.

\section{Evaluation}
\label{eval}
\subsection{Computing Setup}

We evaluate \striderlm on an Amazon EC2/EMR cluster of 11 machines (type m3.xlarge) and manage resources with Yarn. Each machine has 4 CPU virtual cores of 2.6 GHz Intel Xeon E5-2670, 15 GB RAM, 80 GB SSD, and 500 MB/s bandwidth. The cluster consists of 2 nodes for data flow management via the Kafka broker (version 0.8.x) and Zookeeper (version 3.5.x)\cite{Zookeeper}, 9 nodes for Spark cluster (1 master, 8 workers, 16 executors). We use Apache Spark 2.0.2, Scala 2.11.7 and Java 8 in our experiment. The number of partitions for message topic is 16, generated stream rate is around 200000 triples/second.

\subsection{Datasets, Queries and Performance metrics}
Two characteristics prevent us from using well-established RSP benchmarks\cite{SRBench, LSBench, city}: their lack of support for the considered reasoning tasks and their inability to cope with massive RDF streams. We thus created a stream generator based on the Lehigh University Benchmark (henceforth LUBM)\cite{lubm} with 10 universities, \ie containing 1.4 million triples. For the purpose of our experimentation, we extended LUBM with triples containing the \sameAs property. This extension requires to set two parameters: the number of cliques in a dataset and the number of distinct individuals per clique. To define these parameters realistically, we ran an evaluation over different LOD datasets. The results are presented in Table \ref{table:sameAs}. It highlights that although the number of cliques can be very large (over a million in DBpedia), the number of individuals per clique is rather low, \ie a couple of individuals. Given the size of our dataset, we will run most of our experimentations with 1.000 cliques and an average of 10 individuals per clique, denoted 1K-10. Nevertheless, on queries requiring this form of reasoning, we will stress \striderlm with up to 5.000 cliques and an average of 100 individuals per clique (see Fig.\ref{fig:q7-q8} for more details). Figure \ref{table:matTrip} presents the impact on the size of materialized triples for different clique configurations.

We have defined a set of 8 queries\footnote{https://github.com/renxiangnan/strider/wiki} to run our evaluation (see Appendix for details). Queries Q1 to Q5 are limited to concept or/and property subsumption reasoning tasks. Query Q6 implies sameAs only inferences while Q7 and Q8 mix subsumptions and sameAs inferences.

Finally, we need to define which dimensions we want to evaluate. According to \emph{Benchmarking Streaming Computation Engines at Yahoo!}\footnote{https://yahooeng.tumblr.com/post/135321837876/benchmarking-streaming-computation-engines-at}, a recent benchmark for modern distributed stream processing framework,  we take system throughput and query latency as two performance metrics. In this paper, throughput refers to how many triples can be processed in a unit of time (\eg triples per second). Latency indicates the time consumed by an RSP engine between the arrival of the input and the generation of its output.

\subsection{SameAs Materialization - SAM}
Unfortunately, we can not compare \striderlm to other available RSP systems. Since the stream rate generated in the experiment is extremely high, the state-of-the-art reasoning-enabled RSPs like C-SPARQL and SparqlStream behave abnormally. This is probably caused by the scalability of data flow management in such RSPs. Overall, the design of these RSP engines was not intended for large-scale streaming data processing. 
For these reasons, we have designed a Spark streaming engine that integrates a  materialization approach. A full materialization approach, \ie generating Fig.\ref{fig:sameAsRep}(b) from Fig.\ref{fig:sameAsRep}(a), is too costly and naïve in a streaming context. Hence we did not considered it. In fact, we defined a lighter system that is denoted SameAs Materialization (henceforth SAM) where only bi-directional \sameAs triples between sameAs individuals in the data streams are generated. In Fig.\ref{fig:sameAsRep}(b), it would correspond to create the bi-directional red dotted edges and to make all sameAs triples bi-directional, \ie a triple $a~\sameAs~b$ implies $b~\sameAs~a$. SAM also comes with a peculiar query rewriting\footnote{Available on our github page} corresponding to the renaming of BGP  variables with unique variable names, \ie $?x~type~C. ?x~hasName~A$. becomes $?x~type~C. ?y~hasName~A$. Then these new variable names are joined to a common new variable via sameAs triple patterns, \ie the previous BGP is extended with $?x~\sameAs~t. ?y~\sameAs~t.$ In terms of concept and property subumption inferences, SAM adopts the standard query rewriting that introduces UNION clauses between combinations of BGP reformulation. Such a rewriting comes at the cost of increasing the number of joins.
 Table \ref{table:joins} sums up the join and union operations involved in the 8 queries of our experimentation. 
 In particular, queries Q5, Q7 and Q8 present an important number of joins (resp. 90, 45 and 180) due to a large number of union clauses (resp. 17, 14, 29). Appendix F provides some implementation details on the materialization of sameAs individuals in SAM.


\begin{table}[h]
  \captionsetup{justification=centering}
\advance\leftskip-0.15cm
\begin{tabular}{|l|c|c|c|c|c|c|c|c|}
\hline
Cliques  & 1k-10 & 1k-25 & 1k-50 & 1k-100 & 2k-10 & 5k-10  \\
\hline
SAM & $10^5$ & $6,25 . 10^5$ & $2,5 . 10^6$ & $10^7$ & $2.10^5$ & $5.10^5$ \\
\hline
\end{tabular}
\caption{Number of Materialized triples in SAM; \\ 
1K-10 signifies 1000 cliques, 10 individuals per clique }
\label{table:matTrip}
\vspace{-3mm}
\end{table}

\begin{table}[h]
\begin{tabular}{|l|c|c|c|c|}
\hline
datasets &
\#triples & $\vert$sameAs cliques$\vert$ & max(ipc) & avg(ipc) \\
\hline
\hline
DBPedia* & 1032723 & 1032723 & 2 & 2 \\
\hline
Yago* &3696623 & 3696622 & 2 & 2 \\
\hline
Drugbank & 4215954 & 7678 & 2 & 2 \\
\hline
Biomodels & 2650964 & 187764 & 2 & 1.95 \\
\hline
SGD & 14617696 & 15235 & 8 & 3 \\
\hline
OMIM & 9496062 & 22392 & 2 & 2 \\
\hline
\end{tabular}
\caption{SamesAs statistics on LOD datasets (ipc = number of distinct individuals per sameAs clique, *: subsets containing only \sameAs triples, Biomodels contains triples of the form $a~\sameAs~a$}
\label{table:sameAs}
\vspace{-3mm}
\end{table} 

The window size for involved continuous SPARQL queries with LiteMat reasoning support is set to 10 seconds, which is large enough to hold all the data generated from the dataset. However, since the impacts of extra data volume and more complex overheads are introduced in SAM query processing, we have to increase the window size (up to 60 seconds) to ensure that both LiteMat and SAM approaches return the same result. In a nutshell, we approximately adjust the window size and the incoming stream rate by checking the materialized data volume.

\begin{table}[h]
\begin{tabular}{|l|c|c|c|c|c|c|c|c|}
\hline
\hline
Queries & Q1 & Q2 & Q3 & Q4 & Q5 & Q6 & Q7 & Q8 \\
\hline
\#joins LM& 1 & 4 & 0 & 2 & 5 & 2 & 3 & 5 \\
\hline
\#joins SAM & 3 & 24 & 0 & 18 & 90 & 5 & 45 & 180 \\
\hline
\#union LM & 0 & 0 & 0 & 0 & 0 & 0 & 0 & 0 \\
\hline
\#union SAM & 2 & 5 & 2 & 8 & 17 & 1 & 14 & 29 \\
\hline
\end{tabular}
\caption{Number of joins per query for LiteMat and SAM (query rewriting) approaches}
\label{table:joins}
\vspace{-3mm}
\end{table}


\subsection{Results evaluation \& Discussion}

\begin{figure*}[h]
  \centering
  \subfigure[Throughput Comparison for \textbf{Q1} to \textbf{Q5}]{\includegraphics[keepaspectratio=true,scale=0.45]{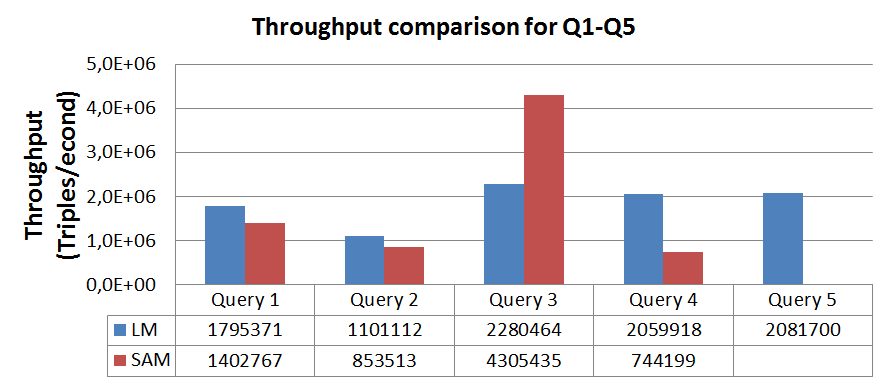}}
  \subfigure[Latency Comparison for \textbf{Q1} to \textbf{Q5}]{\includegraphics[keepaspectratio=true,scale=0.45]{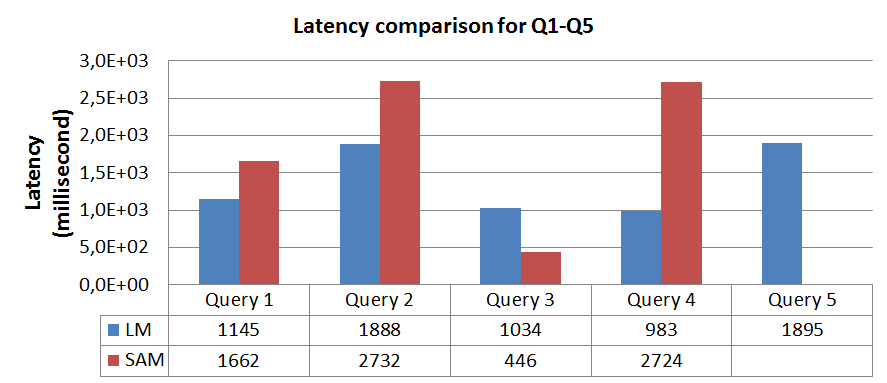}}
\caption{Throughput, Latency Comparison between LiteMat and SAM for \textbf{Q1} to \textbf{Q5}}
\label{fig:q1-q5}
\vspace*{-3mm}
\end{figure*}

\begin{figure*}[h]
  \centering
  \captionsetup{justification=centering}
  \subfigure[Throughput Comparison for \textbf{Q6}]{\includegraphics[scale=0.52]{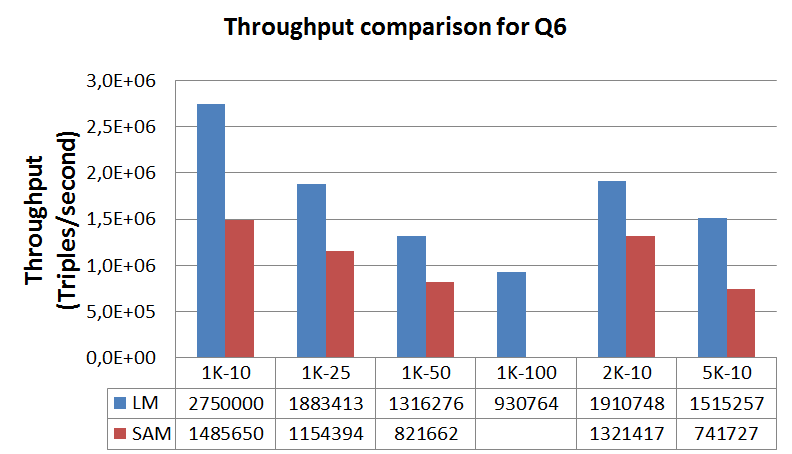}}\hfil
  \subfigure[Latency Comparison for \textbf{Q6}]{\includegraphics[scale=0.52]{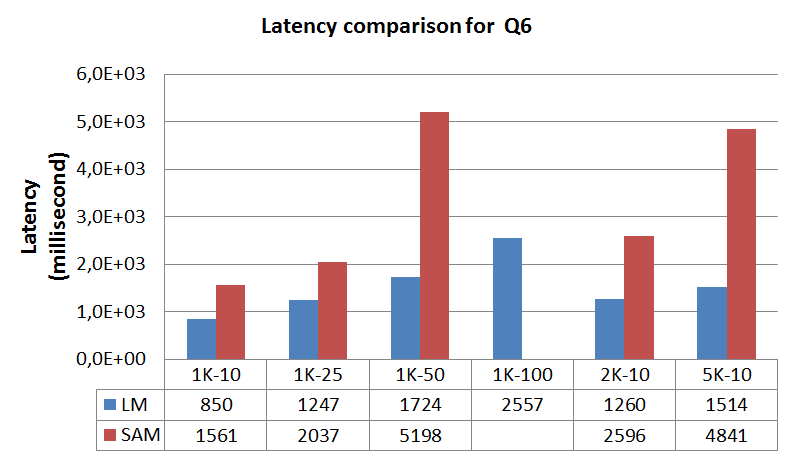}}
\caption{Throughput, Latency Comparison between LiteMat (LM) and SAM for \textbf{Q6} by varying the size of clique.}
\label{fig:q6}
\vspace*{-3mm}
\end{figure*}

\begin{figure*}[!h]
  \centering
  \captionsetup{justification=centering}
  \subfigure[Throughput Comparison for \textbf{Q7} and \textbf{Q8}]{\includegraphics[scale=0.52]{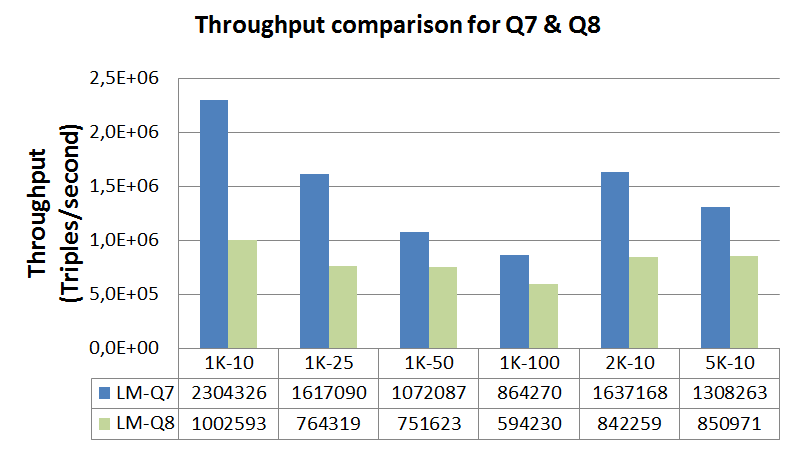}}\hfil
  \subfigure[Latency Comparison for \textbf{Q7} and \textbf{Q8}]{\includegraphics[scale=0.52]{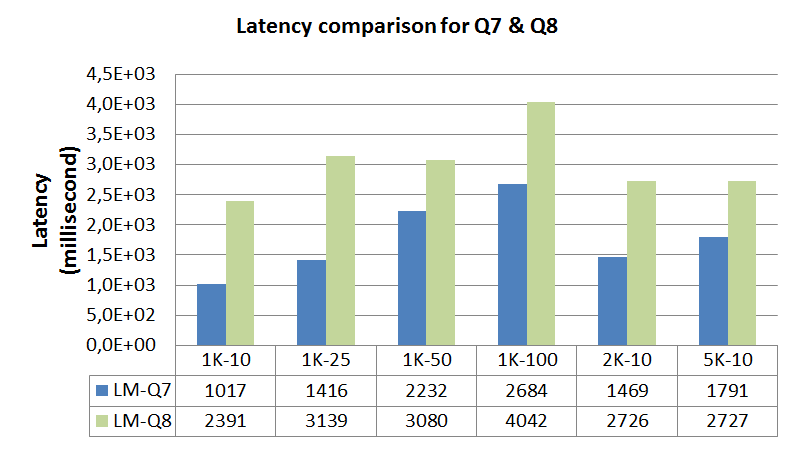}}
 \caption{Throughput, Latency Comparison of LiteMat (LM) for \textbf{Q7} and \textbf{Q8} by varying the size of clique.}
\label{fig:q7-q8}
\end{figure*}

All the evaluation results include the cost of LM encoding and SAM stream materialization. Figure \ref{fig:q1-q5} reports the throughput and query latency of Q1 to Q5. When reasoning with LM is triggered, \striderlm achieves million-level throughput (up to 2 millions triples/seconds), and the query latency remains at a second-level. To the best of our knowledge, such performances have not been achieved by any existing RSP engines. When both original and rewritten query shapes are relatively simple, \eg Q1 and Q2, LM has thirty percent gain over SAM on throughput and latency. The improvement gain of LM over SAM is increasing for queries involving  multiple inferences: Q4 is 75\% faster than SAM. For Q5, SAM does not even terminate. This is mainly due to the insufficient computing resources (\eg number of CPU cores, memories) on Spark driver nodes, which is not capable of handling such intensive overheads for jobs/tasks scheduling and memory management. Nevertheless, SAM is more efficient than LM on Q3 which contains a single triple pattern. This is due to the automatic parallelism provided by Spark on the execution of the \texttt{UNION} queries, thus benefiting from a good usage of cluster resources. With its \texttt{FILTER} clause, LM on Q5 does not benefit from such a parallel execution. Under this circumstance, a filter operator with numeric range determination seems to be more costly than the union of three selections. Such a use case motivates our configuration (Section \ref{config}) where the end-user can force a reasoning approach.



Figure \ref{fig:q6} and Figure \ref{fig:q7-q8} illustrate the impact on engine throughput and latency of Q6 to Q8 with varying sameAs clique sizes. As noted previously, ``1K-10" means 1000 cliques, and 10 individuals per clique. 
Table \ref{table:matTrip} summarizes the number of materialized triples for sameAs reasoning support. The number of materialized triples obviously increases with greater number of cliques and/or number of individuals per clique.
The data throughput and latency can only be compared on Q6 since on Q7 and Q8, SAM does not terminate. The same non termination issue than on Q5 is observed (Q7 and Q8 respectively have 45 joins and 180 joins). Although stream rate is controlled at a low level, the system quickly fails after the query execution is triggered.
Data throughput and latency is always better for LM than SAM by up to respectively an order of magnitude of 2.5 to 3. For the same computing setting, when the number of individuals per clique increases for a given number of cliques or when the number of cliques increases for the  same number of individuals per clique, the performances of the LM approach decreases.  

If LM approach of \striderlm is always better than the SAM approach
On Q6, the latency of SAM substantially increases when the size of the clique is raised to 1K-100, as SAM requires $10^7$ extra triples to be materialized.
The same evolution is witnessed for the more complex Q7 and Q8 queries. Nevertheless, for these queries, throughput of over 800.000 triples per second can be achieved. Given our computing setting of 11 machines, this is still a major breakthrough compared to existing RSP engines which are not able to support important constructors such as \sameAs.


\section{Related work}
This section considers two related work fields: reasoning over ontologies supporting \sameAs and RSP.
Most RDF stores are using a more or less advanced form of encoding and do not adopt a full materialization approach due to its inefficiency. Concerning the support of \sameAs inferences, GraphDB Enterprise Edition \footnote{http://graphdb.ontotext.com/} and RDFox \cite{DBLP:conf/semweb/NenovPMHWB15} are using a representative-based approach but do not handle stream processing. For instance, RDFox elects a representative among elements of a \sameAs clique using a naive lexicographic order. Then all occurrences of individuals of a clique are replaced by the representative. This presents drawbacks when updates are performed on the clique, \eg removing the representative from the clique. In such situations, the original dataset has to be processed again. In the context of a streaming, where data streams are ephemData throughput and latency is always better all, and thus not persisted, update operations on the clique are easily handled. 
\cite{DBLP:conf/semweb/PotterMNH16} follows on the work of RDFox but considers a distributed approach. Nevertheless, the system is not fault tolerant and does not addresses stream processing.
The Kognac system\cite{kognac} proposes an intelligent encoding of RDF terms for large KBs. It is designed on a combination of estimated frequency-based encoding and semantic clustering. Nevertheless, Kognac is not designed with inferences in mind and its implementation is not distributed and thus can not scale to very large KBs.
Laser\cite{laser} is a stream reasoning system based on a tractable fragment of  LARS\cite{lars}, \ie an extension of Answer Set Programming for stream processing). The reasoning services of Laser are supported by a set of rules which are expressed in a specific syntax. We consider that this may prevent Laser's adoption by end-users. Moreover, Laser is not distributed and is thus not able to process very large data streams.


The first RSP engines \cite{CSPARQL,CQELS,SPARQLStream,ETALIS} have emerged around 2009. Their original focus was on the design of continuous query languages based on SPARQL. Scalability and reasoning are now considered as primordial features.  \cite{S4SR} was among the first systems to concentrate on the scalability of RDFS stream reasoning. The engine is able to reach throughputs around hundreds of thousand triples/seconds within 32 computing nodes. However, \cite{S4SR} does not include \sameAs in the scope of consideration. Using standards such Apache Kafka and Spark streaming enabled \striderlm to increase throughput with less computation power by an order of magnitude while being able to reason over RDFS + \sameAs.

\section{Conclusion}
In this paper, we have presented the integration of our LiteMat reasoner for RDFS plus \sameAs within our Strider RSP engine. To the best of our knowledge, this is the first scalable, production-ready RSP system to support such an ontology expressiveness. Via a thorough evaluation, we have demonstrated the pertinence of our system to reason with low latency over high throughput data streams. One of the limitations of our system corresponds to the potential large memory footprint of the generated dictionaries. 
Therefore, as future work, we will introduce an efficient partitioning solution of these dictionaries across a cluster of machines. An improvement of the FILTER  operator in \striderlm is also in the scope of consideration.   
To reach its full potential, this approach will be combined with a partitioning of the streaming data. Finally, we are also working on increasing the expressiveness of supported ontologies, \eg including transitive properties.

\bibliographystyle{abbrv}
\bibliography{main} 

\appendix
In this Appendix section, we provide the eight SPARQL queries evaluated in Section \ref{eval}. In all of them, we are using the \texttt{rdf} and \texttt{lubm} namespaces which respectively correspond to http://www.w3.org/1999/02/22-rdf-syntax-ns\# and http://swat.cse.lehigh.edu/onto/univ-bench.owl\#.

\subsection{Queries with inferences over concept hierarchies}
Q1: Inferences are required on the Professor concept which has no direct instances in LUBM datasets. 

\begin{center}
\small
\begin{verbatim}
SELECT ?n WHERE {
 ?x rdf:type lubm:Professor; lubm:name ?n.}
\end{verbatim}
\end{center}

Q2: Inferences are required on both the Professor and Student concepts.

\begin{center}
\small
\begin{verbatim}
SELECT ?ns ?nx WHERE {
 ?x rdf:type lubm:Professor; lubm:name ?nx.
 ?s lubm:advisor ?x; rdf:type lubm:Student.
 ?s lubm:name ?ns. }
\end{verbatim}
\end{center}

\subsection{Query with inferences over property hierarchies}
Q3: Inferences are required for the memberOf property which has on direct sub property and one indirect sub property.

\begin{center}
\small
\begin{verbatim}
SELECT ?x ?o  WHERE { ?x lubm:memberOf ?o.}
\end{verbatim}
\end{center}

\subsection{Queries with inferences over both concept and property hierarchies}
Q4: This query mixes the Q1 and Q3 and thus necessitates to reason over the Professor and memberOf hierarchies/

\begin{center}
\small
\begin{verbatim}
SELECT ?o ?n WHERE { 
?x rdf:type lubm:Professor; memberOf ?o;
lubm:name ?n.}
\end{verbatim}    
\end{center}

Q5: This query goes further than Q4 by mixing Q2 and Q3, i.e., it requires reasoning over the Professor and Student concept hierarchies and the memberOf property hierarchy.    

\begin{center}
\small
\begin{verbatim}
SELECT ?ns ?nx ?o WHERE {
 ?x rdf:type lubm:Professor; lubm:name ?nx;
 lubm:memberOf ?o.
 ?s lubm:advisor ?x; rdf:type lubm:Student;
 lubm:name ?ns. }
\end{verbatim}
\end{center}

\subsection{Query with inferences over the owl:sameAs property}
Q6: Inferences are required over a clique of similar individuals of the type PostDoc.
\begin{center}
\small
\begin{verbatim}
SELECT ?n ?e WHERE { ?x rdf:type lubm:PostDoc;
 lubm:name ?n; lubm:emailAddress ?e.}
\end{verbatim}
\end{center}

\subsection{Queries with inferences over concept, property hierarchies and owl:sameAs}
Q7: Inferences over the Faculty concept hierarchy, which includes PostDoc sameAs individuals and the memberOf property.

\begin{center}
\small
\begin{verbatim}
SELECT ?o ?n WHERE {
 ?x rdf:type lubm:Faculty; memberOf ?o; 
 lubm:name ?n.}
\end{verbatim}
\end{center}

Q8: The most complex query of our evaluation with two inferences over concept hierarchies (Faculty and Student), with the former containing sameAs individual cliques, and inferences over the memberOf property hierarchy.
\begin{center}
\small
\begin{verbatim}
SELECT ?ns ?nx ?o WHERE { 
 ?x rdf:type lubm:Faculty; lubm:name ?nx; 
 lubm:memberOf ?o. 
 ?s lubm:advisor ?x; rdf:type lubm:Student;
 lubm:name ?ns.}

\end{verbatim}
\end{center}

\subsection{SAM Stream Materialization}
\begin{figure}[ht]
\advance\leftskip-0.3cm
\includegraphics[scale=0.35]{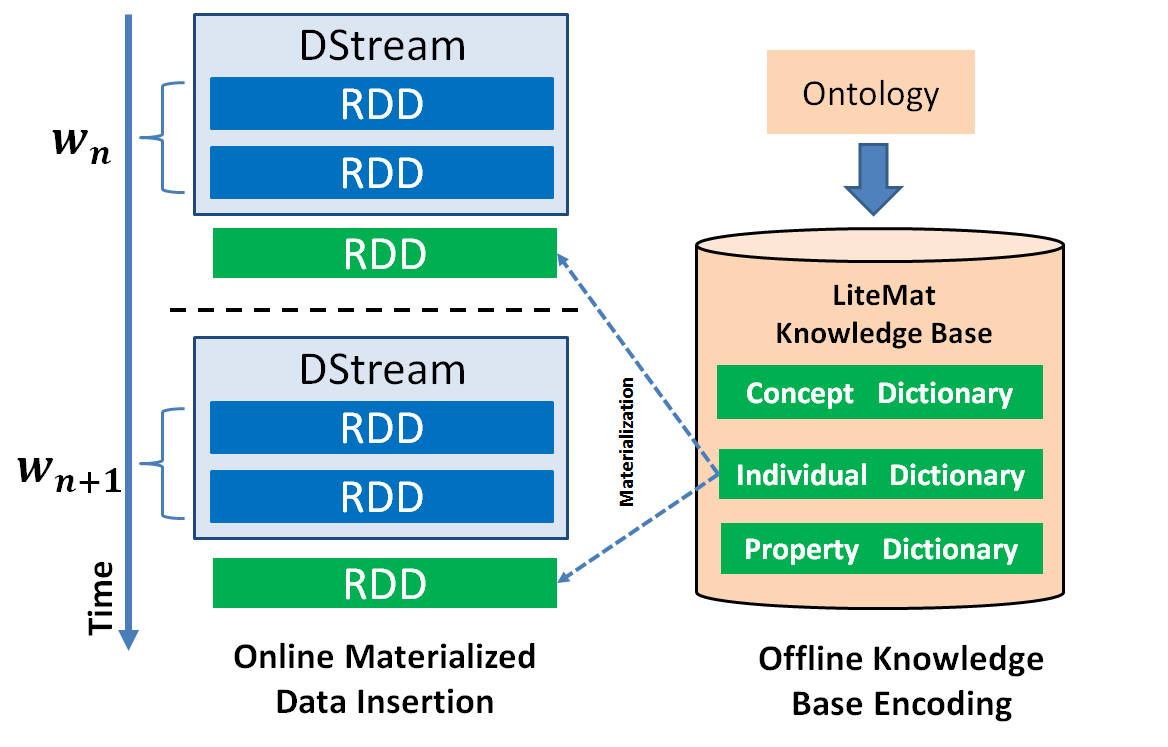}
\caption{SAM stream materialization via LM knowledge base }
\label{fig:lm-stream-mat}
\vspace{-3mm}
\end{figure}

Figure \ref{fig:lm-stream-mat} provides an overview of the SAM stream materialization for reasoning over \sameAs property. 
First, the system parses an input ontology and creates the dictionary of concept/property/individual ($D_{cpt}$/$D_{prop}$/$D_{idv}$).
The dictionary of individual $D_{idv}$ supports the materialization of the extra data $R_{mat}^{SAM}$. This is one improvement that we provide to SAM since parsing the complete KB to detect sameAs triples would be highly inefficient.
When newly stream is injected into the system, we generate the extra materialized data fragment for SAM via a lookup in $D_{idv}$. 
Before the system launches the query processing on newly buffered stream $w_n$, we combine $R_{mat}^{SAM}$ and $w_n$ on the fly, \ie the materialized data stream $w'_n = R_{mat}^{SAM} \cup w_n$. The query evaluation will be executed over $w_n'$ instead of $w_n$. 


\end{document}